\documentclass[12pt,a4paper]{article}

\usepackage{graphicx}

\newlength{\abstwidth}
\newenvironment{Itemize}{\begin{list}{$\bullet$}%
{\setlength{\topsep}{0.2mm}\setlength{\partopsep}{0.2mm}%
\setlength{\itemsep}{0.2mm}\setlength{\parsep}{0.2mm}}}%
{\end{list}}

\begin{document}
\sloppy

\begin{flushright}
LU TP 13-37\\
MCnet-13-16\\
October 2013
\end{flushright}

\vspace{\fill}

\begin{center}
{\LARGE\bf Colour reconnection and its effects\\[4mm] 
on precise measurements at the LHC}\\[10mm]
{\Large Torbj\"orn Sj\"ostrand} \\[3mm]
{\it Theoretical High Energy Physics,}\\[1mm]
{\it Department of Astronomy and Theoretical Physics,}\\[1mm]
{\it Lund University,}\\[1mm]
{\it S\"olvegatan 14A,}\\[1mm]
{\it SE-223 62 Lund, Sweden}
\end{center}

\vspace{\fill}

\begin{center}
\begin{minipage}{\abstwidth}
{\bf Abstract}\\[2ex]
There are experimental evidence for the occurrence of colour reconnection, 
but the mechanisms involved are far from understood. Previous reconnection 
studies are briefly summarized, and some potential implications for 
LHC physics are outlined.
\end{minipage}
\end{center}

\vspace{\fill}
\noindent\rule{60mm}{0.3mm}\\
{\footnotesize To appear in the proceedings of the 
XLIII International Symposium on Multiparticle Dynamics (ISMD13),
Chicago, USA, 15 -- 20 September 2013}

\clearpage

\section{Introduction}

LHC events have a complicated structure, which involves many physics 
components, the main ones being hard-process matrix elements, 
parton distribution functions, multiple parton interactions (MPIs), 
initial-state radiation (ISR), final-state radiation (FSR), 
beam remnants,  hadronization and decays. All of these contain 
challenges, but are still understood individually, to some extent. 
When combined, additional sources of uncertainty appear, 
however. Foremost among these, colour reconnection (CR) represents
the uncertainty induced by the high density of colour charges, that
may interact in a nontrivial nonlinear manner.

To put numbers on the challenge, about ten charged particles are
produced per unit of rapidity for LHC events at around $y = 0$.
These come from around ten primary hadrons, 
which in their turn come from ten colour strings
\cite{lundString} crossing $y = 0$, according to \textsc{Pythia} 
\cite{pythia8} simulations. The distributions are very widely
spread around this average, so much higher densities are common.
The string density is largely driven by the MPI component, where
each gluon--gluon scattering may lead to two strings crossing  $y = 0$,
but it also receives contributions from ISR and FSR. The string width 
is the same as that of a hadron, the two being dictated by the same
confinement physics, and most of the strings are produced and evolve
within the transverse area of the original proton--proton collision. 
Therefore many strings overlap in space and time, potentially leading 
to nonlinear effects. Furthermore, the small number of colours, 
$N_C = 3$, inherently leads to ambiguities which partons belong
together in separate colour singlets.

One approach to this issue would be modify or abandon existing 
hadroni\-zation models, colour ropes \cite{ropes} being an example 
of the former and quark--gluon plasma of the latter. Less dramatic 
is the CR road, where hadronization as such is unmodified and the 
nonlinear effects are introduced via models that ``only'' reassign 
colours among partons. In the following we will study such models 
and some of their consequences. 

\section{Historical overview}

The idea that colour assignments provided by perturbation theory could 
be modified by nonperturbative effects was around already soon after 
the birth of QCD. The colour octet production mechanism 
$g^* \to c\overline{c} \to J/\psi$ \cite{octetFritzsch} is an early 
example. Such colour rearrangement effects were studied more 
systematically for $B$ decay \cite{reconAli}, 
and the sequence $B \to J/\psi \to \ell^+\ell^-$ was proposed as an 
especially convenient test \cite{reconFritzsch}. Indeed the 
$B \to J/\psi$ branching ratio suggests a non-negligible but not 
dominant fraction of the $b \to c W^- \to c \overline{c} s$ rate, 
kinematical restrictions taken into account \cite{reconCC}.    

Colour reconnection in minimum-bias hadronic physics was first 
introduced \cite{mpiZijl} to explain the rising trend of 
$\langle p_{\perp} \rangle (n_{\mathrm{ch}})$ observed by UA1
\cite{ptnchUA1}. The starting point here is that large charged-particle 
multiplicities predominantly come from having a large MPI activity, 
rather than from high-$p_{\perp}$ jets, say. If each such MPI produces
particles more-or-less independently of each other, then the 
$\langle p_{\perp} \rangle$ should be independent of the number of MPIs, 
and hence of $n_{\mathrm{ch}}$. The alternative is that each further
MPI brings less and less additional $n_{\mathrm{ch}}$, while 
still providing an equally big $p_{\perp}$ kick from the (semi-)hard 
interaction itself, to be shared among the produced hadrons. This is 
possible in scenarios with CR, if reconnections tend to reduce the 
total string length $\lambda$ \cite{lambdaMeasure},
\begin{equation}
\lambda \approx \sum_{i,j} \ln\left( \frac{m_{ij}^2}{m_0^2} \right) ~,
\label{lambda}
\end{equation} 
where $i,j$ runs over all colour-string-connected parton pairs and 
$m_0 \approx 1$~GeV is a reference scale of a typical hadronic mass.

As an aside, other aspects (well modeled in generators) drive the 
rise of $\langle p_{\perp} \rangle (n_{\mathrm{ch}})$ at small
$n_{\mathrm{ch}}$. Furthermore, the absolute normalization of 
$\langle p_{\perp} \rangle$ in this region comes straight from tunes 
of hadronization to $e^+e^-$ data, supporting the notion that 
beam-remnant hadronization is no different from that of jets 
so long as the string density is low.    

$W$ pair production at LEP~2 was expected to offer an interesting test 
bed for such concepts, i.e. whether the $q\overline{q}$ pair 
produced in each $W$ decay would hadronize separately or whether 
e.g.\ the $q$ from one $W$ could hadronize together with the 
$\overline{q}$ from the other. Notably, this could mess up $W$ mass 
determinations. Unfortunately, results were not conclusive.
\begin{Itemize} 
\item Perturbative effects are suppressed for a number of reasons,
notably that hard-gluon exchanges would force the $W$ propagators
off-shell, giving a negligible uncertainty 
$\langle \delta M_W \rangle \leq 5$~MeV \cite{reconKhoze}.
\item Several nonperturbative CR models predicted large effects and 
could promptly be ruled out. More conservative ones \cite{reconKhoze} 
could not be excluded, but were not favoured \cite{reconData}, 
and gave $\langle \delta M_W \rangle \sim 40$~MeV.
\item Additionally Bose-Einstein effects, i.e.\ that the wave function
of identical integer-spin hadrons should be symmetrized, could 
affect the separate identities of the $W^+$ and $W^-$ decay products.
Effects on $\langle \delta M_W \rangle$ could be as large as
100~MeV, but again more likely around 40~MeV \cite{beLonnblad}. 
An effect of the latter magnitude is disfavoured by data, but again 
not fully ruled out \cite{beData}. 
\end{Itemize}

Given the clean LEP environment, it was feasible to trace the 
space--time evolution of the strings \cite{reconKhoze}, and use that 
to decide if and where a reconnection would occur. Two alternative 
scenarios were inspired by Type II and Type I superconductors. 
In the former, narrow vortex lines at the core of the strings carry 
the topological information, and so it was assumed that strings could 
reconnect only if and where these cores crossed. In the latter, strings
are viewed as elongated bags with no marked internal structure, and  
therefore the reconnection probability was related to the integrated 
space--time overlap of these bags. In both cases reconnections that
reduced the total string length could be favoured. 

A future high-luminosity $e^+ e^-$ collider for the study of Higgs
production would, as a by-product, provide much larger $W^+W^-$ 
samples and thereby allow more precise tests. Assuming an effect is
found, its energy and angular-orientation dependence could
constrain the range of allowed models \cite{lincolKhoze}.

The observation of diffractive event topologies in Deeply Inelastic 
Scattering at HERA has also been interpreted as a consequence of CR
\cite{reconHERA}. This offers an alternative to the Ingelman--Schlein
picture \cite{IngSch} of scattering on a Pomeron (or glueball, in 
modern language) component inside the proton. Both approaches can be
tuned to give comparable phenomenology, so there is no clear winner
at HERA. Nevertheless, HERA, Tevatron and LHC diffractive data can 
provide significant constraints on any universal model of colour 
reconnection. This also includes topics such as diffractive jet, 
$W$ and Higgs production. Diffraction and models for diffraction 
is such a major topic in its own right \cite{reviewIng} that it is
impossible to cover it here.

It is also plausible that both CR and Pomeron mechanisms contribute 
to the appearance of rapidity gaps. To exemplify, a rapidity gap 
between two high-$p_{\perp}$ jets likely is 
dominated by reconnection, whereas small-mass diffraction comes 
more naturally in a traditional Pomeron language. 

\begin{figure}
\centering
\includegraphics[width=0.6\columnwidth]{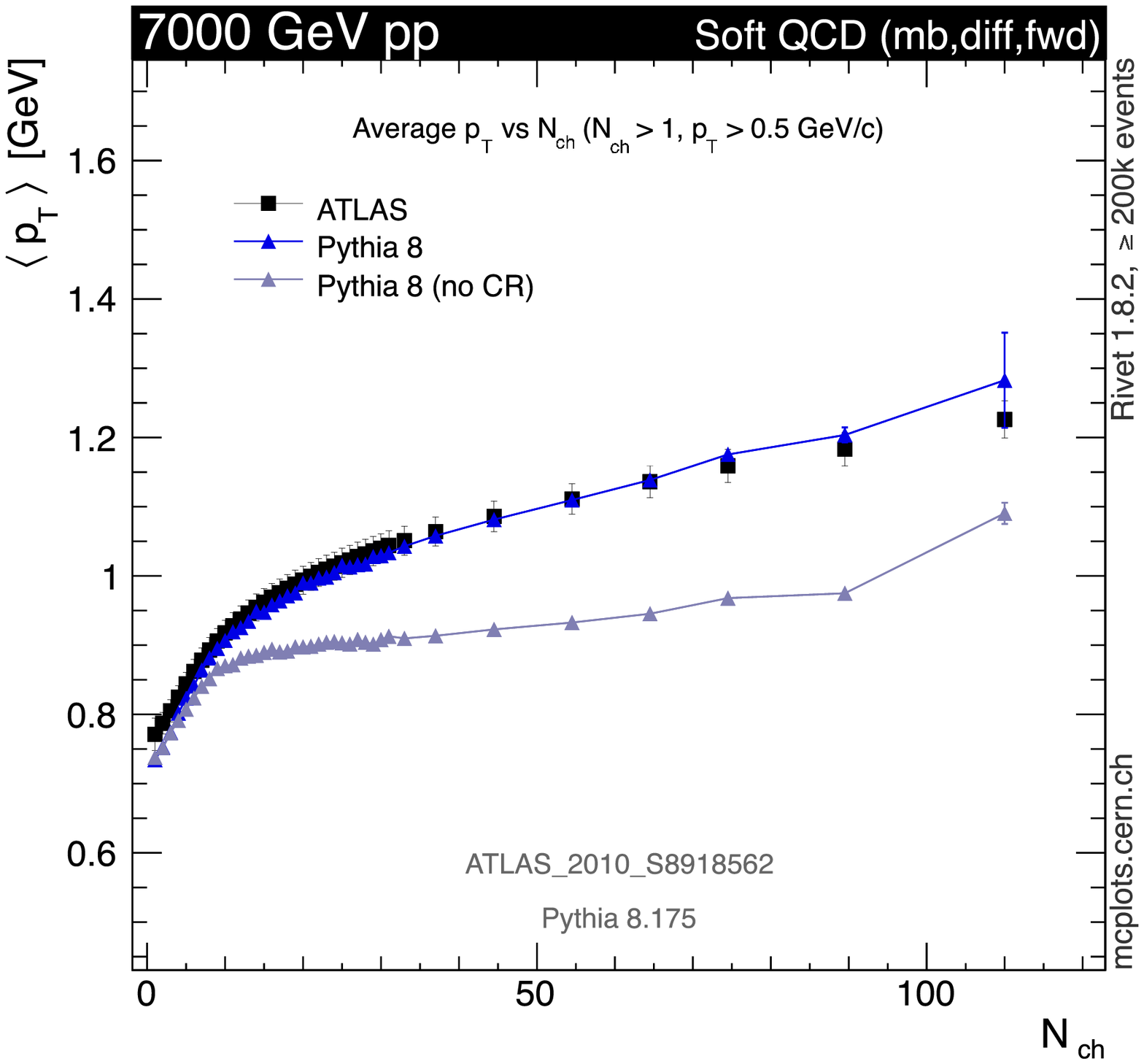}
\caption{$\langle p_{\perp} \rangle (n_{\mathrm{ch}})$ with the
default \textsc{Pythia}~8 Tune 4C \cite{tuneCorke}, and the same
with CR switched off, compared with ATLAS data \cite{pTvsNchATLAS}.  
\label{figPTNCH}}
\end{figure}

\section{Status at the LHC}

While most of the basic ideas for MPI modelling existed a long time 
ago \cite{mpiZijl}, gradually models have become more sophisticated. 
One key example is the handling of beam remnants \cite{brSkands}. 
As a starting point, the colour flow in each separate MPI, including 
its associated ISR and FSR, is traced in the $N_C \to \infty$ 
limit \cite{planarQCD}. (This limit gives a well-defined colour 
topology, as needed for the string  hadronization model.) But any 
colour coming into an MPI must be compensated by a corresponding 
anticolour left behind in a remnant, which for $N_C \to \infty$ 
leads to a remnant momentum to be shared between a multitude of 
string endpoints. Such a scenario is not ruled out, since essentially 
no data exists on how the remnant structure changes as a function of 
the central multiplicity, and since a modelling could introduce many 
free tuning parameters, but neither is it plausible.  

Instead it is likely that the $N_c = 3$ reality leads to a smaller
remnant colour charge, as the initial colour of one MPI often 
compensates the anticolour of another, thereby correlating the 
colour flow of these two MPIs right up to the final state.  
Such correlations means that fewer strings need to be drawn out 
to the beam remnants for high MPI multiplicities, offering a 
mechanism for a rising $\langle p_{\perp} \rangle (n_{\mathrm{ch}})$,
but nowhere near enough. Thus, also with modern models, LHC data 
reconfirm the need for a further mechanism, such as CR. 
This is illustrated in Fig.~\ref{figPTNCH}. 

The almost perfect agreement in Fig.~\ref{figPTNCH} is fortuitous, 
and it looks less impressive with other selection criteria
\cite{pTvsNchATLAS,pTvsNchCMS}, even if 
the qualitative features still are reproduced. So there is room for 
improvements of the CR modelling, or for other physics mechanisms. 

Over the years, \textsc{Pythia}~6.4 has come to contain a dozen of 
CR scenarios, many closely related. Unlike the above-mentioned 
$e^+e^-$ scenarios there is no attempt to trace a space--time 
evolution. Instead the guiding principle is to reduce the total 
string length, as defined by the $\lambda$ measure of 
eq.~(\ref{lambda}) or, alternatively, by the $\sum_{i,j} m_{ij}^2$ 
(GAL, Generalized Area Law \cite{GALrathsman}). Typically an 
algorithm may go something like \cite{topSkands}
\begin{Itemize}
\item Calculate a reconnection probability 
$P_{\mathrm{rec}} = 1 - (1 - \chi)^{n_{\mathrm{MPI}}}$,
where $n_{\mathrm{MPI}}$ is the number of MPIs in the current event 
and $\chi$ is a free reconnection strength parameter.
\item Each string piece is chosen to be a candidate for reconnection 
with a probability $P_{\mathrm{rec}}$.
\item Use a simulated annealing algorithm to perform reconnections
between the candidates picked in the previous step, favouring a
reduced $\lambda$.
\end{Itemize}

By contrast, currently \textsc{Pythia}~8.1 only contains one scenario,
where either all or none of the final-state partons of a MPI system
are attached to the string pieces of a higher-$p_{\perp}$ system,
in a way so as to keep $\lambda$ minimal. The lower the $p_{\perp}$
scale of an MPI, and the larger the number of other MPIs, the more 
likely it is to be disassembled by CR.   

Also the other standard LHC generators face similar issues.
The inclusion of CR into \textsc{Herwig/Herwig++} \cite{HerwigMan} is of 
fairly recent date \cite{HerwigRecon}. CR is necessary not only to 
to describe $\langle p_{\perp} \rangle (n_{\mathrm{ch}})$ but also e.g.\
the $\mathrm{d}n_{\mathrm{ch}} / \mathrm{d}\eta$ distribution. 
Again a simulated annealing approach is used to reduce $\sum m^2$,
where the sum runs over all clusters, akin to the GAL above. 
\textsc{Sherpa} \cite{SherpaMan} currently has an MPI model based on 
the \textsc{Pythia} one, but without any colour reconnection.   
Therefore it also fails to describe the 
$\langle p_{\perp} \rangle (n_{\mathrm{ch}})$ distribution.
A new model for minimum-bias and underlying events is in preparation
\cite{SherpaRecon} that should address it.

\section{The mass of unstable coloured particles}

Confinement leads to ambiguous masses for coloured particles,
since they cannot be studied in isolation. Short-lived coloured
particles, like the top, do not even form hadrons with well-defined 
masses. For the kinematics of production and decay, an event generator 
therefore have to use its own mass definition, that is close to 
but not necessarily identical with the pole mass. This inherently 
leads to ambiguities in a translation  of a generator-assisted top
mass measurement into a corresponding $\overline{\mathrm{MS}}$ mass.

Furthermore the top quark, as well as the $W$ and $Z$ gauge bosons, 
travel a distance $c \tau \approx 0.1$~fm before they decay, i.e.\ 
significantly less than a proton radius. Therefore their decays take 
place right in the middle of the showering/hadronization region, and 
so quarks (and gluons) produced in the decays are subject to the CR 
issues already discussed. That is, in a decay $t \to b u \overline{d}$ 
the $b$ for sure is colour-connected somewhere else, giving mass 
ambiguities, but additionally the $u \overline{d}$ system may or may 
not remain as a separate singlet, further contributing to the 
uncertainty. 

Studies with \textsc{Pythia}~6.4 for the Tevatron suggested a total 
uncertainty approaching 1~GeV \cite{topSkands} when comparing different 
tunes. Of this a large part comes from the description of the 
perturbative stage, i.e. ISR and FSR uncertainties, which should 
have shrunk considerably since, with the advent of more sophisticated
matching/merging techniques. But up to 0.5~GeV remains as a potential
error related to CR issues. To put this in context, current top mass 
measurements at the Tevatron and the LHC now have statistical errors 
of the order 0.5 GeV, and quote systematic errors below 1~GeV 
\cite{topData}. 

Clearly this issue needs to be studied further, to try to constrain 
the possible magnitude of effects from data itself. CR effects should 
depend on the event kinematics, which would allow to test and 
constrain models. Such studies have already begun in CMS 
\cite{topStudCMS}, although statistics does not yet allow any 
conclusions to be drawn. 

As already mentioned, \textsc{Pythia}~8.1 does not yet have a range
of CR scenarios to contrast, but CR on or off gives a shift of 
$\approx 0.15$~GeV. Unfortunately this difference does not vary 
dramatically as a function of some obvious kinematical variables, 
but further studies are planned. 

In top decays to leptons, $t \to b \ell^+ \nu_{\ell}$, the lepton 
$p_{\perp}$ spectrum offers a CR-independent observable, that may 
allow an alternative route. It will face other challenges, however. 

\section{Summary and outlook}

Colour reconnection as such is well established, e.g.\ from 
$B \to J/\psi$. Given the high string and particle densities 
involved in a high-energy $pp$ collision, it is hard to imagine
that it would \textit{not} play a prominent role also there. 

This does not mean that what we today ascribe to CR could not
be a much richer mixture of high-density effects, such as colour ropes 
or collective flow. The particle composition as a function of
$p_{\perp}$ is one example of LHC distributions not well described
by \textsc{Pythia} simulations, and where thus some further 
mechanism may be at play. There is a twist to this story, 
however, in that CR in $pp$ events can give some of the observed 
effects similar to the collective flow of heavy-ion collisions 
\cite{collectflow}, by a combination of two factors. Firstly, 
a string piece moving with some transverse velocity tends to 
transfer that velocity to the particles produced from it, albeit 
with large fluctuations, thereby giving larger transverse momenta 
to heavier hadrons. Secondly, a string piece has a larger 
transverse velocity the closer to each other the two endpoint 
partons are moving, which is precisely what is favoured by CR 
scenarios intended to reduce the string length.

In the near future, the intention is to implement new CR models
for $pp$ collisions into \textsc{Pythia}~8, partly to offer a broader 
spectrum of possibilities, partly to add further physics aspects,
such as the space--time and colour structure, to provide more 
realistic scenarios. Other generator authors will also offer their
schemes. When systematically confronted with a broad spectrum of 
data the hope is to see a pattern emerge, where some approaches
are more favoured than others. It would be foolish to promise 
that a unique answer will be found, however; we will have 
to live with CR uncertainties in many precision measurements. The 
top mass is the obvious example, but others are likely to emerge 
as LHC exploration continues. 

In the far future, a high-luminosity $e^+e^-$ Higgs factory 
would offer a second chance to study CR and related effects 
in $W^+W^-$ events.

\section*{Acknowledgments}
Work supported in part by the Swedish Research Council, contract number
621-2010-3326, and in part by the MCnetITN FP7 Marie Curie Initial 
Training Network, contract PITN-GA-2012-315877. Thanks to A.~Karneyeu
for providing Fig.~\ref{figPTNCH}, generated with MCPLOTS/Rivet
\cite{mcplotsRivet}.


\begin{thebibliography}{99}

\bibitem{lundString} 
B.~Andersson, G.~Gustafson, G.~Ingelman and T.~Sj\"ostrand,
Phys.\ Rept.\  {\bf 97}, 31 (1983).

\bibitem{pythia8} 
T.~Sj\"ostrand, S.~Mrenna and P.~Z.~Skands,  
JHEP {\bf 0605}, 026 (2006);
Comput.\ Phys.\ Commun.\  {\bf 178}, 852 (2008).

\bibitem{ropes}	
T.S.~Biro, H.B.~Nielsen and J.~Knoll, Nucl.\ Phys.\ {\bf B245}, 449 (1984).

\bibitem{octetFritzsch}
H. Fritzsch, Phys.\ Lett.\ {\bf B67} 164, 217 (1977).

\bibitem{reconAli}
A. Ali, J.G. K\"orner, G. Kramer, J. Willrodt, 
Z.\ Phys.\ {\bf C1}, 269 (1979).

\bibitem{reconFritzsch}
H. Fritzsch, Phys.\ Lett.\ {\bf B86} 164, 343 (1979).

\bibitem{reconCC}
D.~Eriksson, G.~Ingelman and J.~Rathsman,
Phys.\ Rev.\ {\bf D79}, 014011 (2009).

\bibitem{mpiZijl}
T. Sj\"ostrand and M. van Zijl, Phys.\ Rev.\ {\bf D36}, 2019 (1987).

\bibitem{ptnchUA1}
UA1 Collaboration, Nucl.\ Phys.\ {\bf B335}, 261 (1990).

\bibitem{lambdaMeasure}
B.~Andersson, G.~Gustafson and B.~S\"oderberg, 
Nucl.\ Phys.\ {\bf B264}, 29 (1986).

\bibitem{reconKhoze}
T. Sj\"ostrand and V.A. Khoze, Z.\ Phys.\ {\bf C62}, 281 (1994).

\bibitem{reconData}
L3 Collaboration, Phys.\ Lett.\ {\bf B561}, 202 (2003);\\
OPAL Collaboration, Eur.\ Phys.\ J.\ {\bf C45}, 291 (2006);\\
ALEPH Collaboration, Eur.\ Phys.\ J.\ {\bf C47}, 309 (2006);\\
DELPHI Collaboration, Eur.\ Phys.\ J.\ {\bf C51}, 249 (2007).

\bibitem{beLonnblad}
L. L\"onnblad and T. Sj\"ostrand, Phys.\ Lett.\ {\bf B351}, 293 (1995);
Eur.\ Phys.\ J.\ {\bf C2}, 165 (1998).

\bibitem{beData}
L3 Collaboration, Phys.\ Lett.\ {\bf B547}, 139 (2002);\\
OPAL Collaboration, Eur.\ Phys.\ J.\ {\bf C36}, 297 (2004);\\
ALEPH Collaboration, Phys.\ Lett.\ {\bf B606}, 265 (2005);\\
DELPHI Collaboration, Eur.\ Phys.\ J.\ {\bf C44}, 161 (2005).

\bibitem{lincolKhoze}
V.A.~Khoze and T.~Sj\"ostrand,
Eur.\ Phys.\ J.\ direct {\bf C2}, 1 (2000).

\bibitem{reconHERA}
W.~Buchmuller and A.~Hebecker, Phys.\ Lett.\ {\bf B355}, 573 (1995);\\
A.~Edin, G.~Ingelman and J.~Rathsman, Phys.\ Lett.\ {\bf B366}, 371 (1996).

\bibitem{IngSch}
G.~Ingelman and P.E.~Schlein, Phys.\ Lett.\ {\bf B152}, 256 (1985).

\bibitem{reviewIng}
G.~Ingelman, Int.\ J.\ Mod.\ Phys.\ {\bf A21}, 1805 (2006).

\bibitem{brSkands}
T.~Sj\"ostrand and P.~Z.~Skands, JHEP {\bf 0403}, 053 (2004).

\bibitem{planarQCD}
G. 't Hooft, Nucl. Phys. {\bf B72}, 461 (1974).

\bibitem{tuneCorke}
R.~Corke and T.~Sj\"ostrand, JHEP {\bf 1103}, 032 (2011).

\bibitem{pTvsNchATLAS}
ATLAS Collaboration, New J.\ Phys.\  {\bf 13}, 053033 (2011).

\bibitem{pTvsNchCMS}
CMS Collaboration, arXiv:1310.4554 [hep-ex].

\bibitem{GALrathsman}  
J.~Rathsman, Phys.\ Lett.\ {\bf B452}, 364 (1999).

\bibitem{topSkands}
M.~Sandhoff and P.~Z.~Skands, FERMILAB-CONF-05-518-T;\\
P.~Z.~Skands and D.~Wicke, Eur.\ Phys.\ J.\ {\bf C52}, 133 (2007);\\
D.~Wicke and P.~Z.~Skands, Nuovo Cim.\ {\bf B123}, S1 (2008).

\bibitem{HerwigMan}
M.~B\"ahr {\it et al.}, Eur.\ Phys.\ J.\ {\bf C58}, 639 (2008).

\bibitem{HerwigRecon}
S.~Gieseke, C.~R\"ohr and A.~Siodmok, 
Eur.\ Phys.\ J.\ {\bf C72}, 2225 (2012).

\bibitem{SherpaMan}
T.~Gleisberg {\it et al.}, JHEP {\bf 0902}, 007 (2009).

\bibitem{SherpaRecon}
V.A.~Khoze, F.~Krauss, A.D.~Martin, M.G.~Ryskin and K.C.~Zapp,
Eur.\ Phys.\ J.\ {\bf C69}, 85 (2010).

\bibitem{topData}
CDF and DO Collaborations, arXiv:1305.3929 [hep-ex];\\
CMS Collaboration, arXiv:1307.4617 [hep-ex].

\bibitem{topStudCMS}
CMS Collaboration, CMS-PAS-TOP-12-029, CMS-PAS-TOP-13-007.

\bibitem{collectflow}
A. Ortiz Velasquez, P. Christiansen, E. Cuautle Flores, 
I.A. Maldonado Cervantes and G. Pai\'c,
Phys.\ Rev.\ Lett. {\bf 111} (2013) 042001.

\bibitem{mcplotsRivet}
A.~Karneyeu, L.~Mijovic, S.~Prestel and P.~Skands, 
arXiv:1306.3436 [hep-ph];\\
A. Buckley {\it et al.}, arXiv:1003.0694 [hep-ph]. 

\end{thebibliography}
\end{document}